\documentclass[12pt,aps,PRL, paper]{article}
\usepackage{amssymb}

\usepackage{graphicx}
\usepackage{amsmath}

\input{tcilatex}

\begin{document}

\begin{center}
{\Large Novel Transport Mechanism for Interacting Electrons }

{\Large in Disordered Systems: }

{\Large Variable-Range Resonant Tunneling}
\end{center}

\bigskip

\begin{center}
{\large S.D. Baranovskii}$^{1}${\large \ and I.S. Shlimak}$^{2}$
\end{center}

$\bigskip $

\begin{center}
$^{1}$Institute for Physical Chemistry and Material Sciences Center,

Philipps-University Marburg, D-35032 Marburg, Germany

$^{2}$Jack and Pearl Resnick Institute of Advanced Technology,

Department of Physics, Bar-Ilan University, Ramat-Gan 52900, Israel

\bigskip
\end{center}

{\small To interpret recent experimental observations of the phononless
hopping conduction, we suggest a novel transport mechanism according to
which the current-carrying single electrons move via quantum resonant
tunneling between localized states brought into resonance by fast electron
hops in their environment.}

\begin{center}
\bigskip
\end{center}

Recent experiments on the quantum Hall effect [1,2] and on the hopping
transport in 2D Si MOSFETs [3] and delta-doped GaAs/AlGaAs heterostructures
[4] show that prefactor so in the exponential temperature dependence of the
hopping conductivity has, under certain conditions, a universal form ($%
e^{2}/h$)$f(T/T_{0})$ with $f$ $\approx $1 at $T\approx T_{0}$. Aleiner et
al. [5] have remarked that such an effect can be due to hopping via
electron-electron scattering, as considered by Fleishman et al. [6], rather
than via conventional electron-phonon scattering. We suggest below an
alternative to this mechanism based on non-Hubbard Coulomb correlations
which can account for the observed effects.

To interpret phononless hopping evidenced by experimental observations, it
is necessary to take into account electron-electron interactions, though
such interactions can also play an important role for phonon-assisted
hopping. The role of non-Hubbard Coulomb correlations for the
low-temperature hopping conduction was first pointed out by Knotek and
Pollak [7]. They considered two types of such correlations which are caused
by interaction between electrons on different sites: sequential correlations
(SC) and electron polarons (EP). In the case of SC, the single-electron
current-determining hops can be assisted by preliminary hops between
surrounding states. These preliminary hops prepare the favorable situation
for the current-carrying transitions of single electrons. In the case of EP,
several electrons hop simultaneously. Such simultaneous multielectron
transitions can have lower activation energies than single-electron hops and
therefore they can be efficient [7]. In contrast to the case of the
sequential transitions, the probability for a simultaneous transition of
several electrons involves the product of the factors exp$\{-2r/a\}$ for
each participating single-electron hop (where $r$ is the hop distance and $a$
is the localization length). Therefore, such transitions can be neglected
when the concentration of localized states is sufficiently low, although for
intermediate doping levels such transitions can be efficient [8]. Pollak and
his coworkers have also emphasized the importance of quantum effects in
disordered systems of localized electrons interacting via their Coulomb
potentials [9,10]. By computer simulation they studied the level spacing
spectra as a measure of the degree of localization, though not directly
addressing the conduction phenomena.

The question most widely discussed related the electron-electron
correlations is whether or not the temperature dependences of the hopping
conductivity obtained for single-electron hopping are altered by taking into
account the many-particle effects. Perez-Garrido et al. [11] studied this
problem by computer simulation and concluded that while both correlation
effects have a significant influence on the hopping conductivity, its
temperature dependence at low T is described quite well by the dependence
predicted by theories of single-electron hopping, although with changed
numerical parameters. In the Coulomb gap regime, the Efros-Shklovskii
formula should be valid [8]

\begin{equation}
\sigma \varpropto \exp \{-(T_{1}/T)^{1/2}\},  \label{Eq.1}
\end{equation}
whereas at higher temperatures, the Mott formula applies:

\begin{equation}
\sigma \varpropto \exp \{-(T_{2}/T)^{1/(d+1)}\},  \label{Eq.2}
\end{equation}
where $d$ is the spatial dimensionality and $T_{1},T_{2}$ are characteristic
temperatures.

So far, most research has been concentrated on the phonon-assisted hopping
processes. Only one paper [6] has been devoted to the temperature dependence
of the hopping conductivity in the case of phononless hopping. Fleishman et
al. [6] have studied EP correlations, i.e., simultaneous transitions of
several electrons in the phononless mode. They succeeded in showing that for
the phononless transport of multielectron excitations, the temperature
dependence of the conductivity is described by Eq.(2) in the Mott regime. It
may well be the case, as proposed by Aleiner et al. [5], that it is this
transport mechanism, which is responsible for the recent observations of the
phononless hopping in 2D systems. Not arguing in favour or against this
hypothesis, we suggest below an alternative mechanism of the phononless
hopping processes which also provides the observed temperature dependencies
of the hopping conductivity. We do not aim at calculation of prefactors and
concentrate our discussion on exponential terms.

We assume that the current-carrying hop between the sites belonging to the
percolation cluster is represented by a single-electron resonant transition,
the resonance being prepared by fast assisting hops in the environment of
these sites. It is a kind of SC with the current-carrying hop provided by
resonant tunneling. At first glance, this looks similar to the picture
recently discussed by Agrinskaya and Kozub [12]. They also considered SC and
they also assumed assisting hops to be fast compared to the current-carrying
hop. However, Agrinskaya and Kozub suggested for the rate $W$ of the
current-carrying hop expression 
\begin{equation}
W\varpropto \exp \{-2r/a-\varepsilon /kT\},  \label{Eq.3}
\end{equation}
where the activation term describes the probability of an assisting hop, and 
$r$ is the distance between the corresponding pair of the current-carrying
sites. The activation energy of the current-carrying hop was assumed to
vanish due to averaging over fast assisting hops [12]. As long as the
assisting hops are due to electron-phonon scattering, the phonon frequency
is inherently present in the rate of current-carrying hops of Agrinskaya and
Kozub [12].We argue below that in our mechanism of resonant transitions, the
rate $W$ of a current-carrying hop does not involve the phonon frequency
even if the assisting hops are due to the interaction with phonons.

Let us consider two neighboring sites $i$ and $j$ belonging to the
percolation cluster, so that an electron transition between them is
necessary for dc transport. We assume that because of fast electron
transitions in the sites surrounding $i$ and $j$, the electron energies $%
\varepsilon _{i}$ and $\varepsilon _{j}$ on the chosen sites fluctuate with
some typical frequency $\omega $ due to the Coulomb potentials of
surrounding electrons. In fact, it is sufficient to consider the
fluctuations of just one of the two energies $\varepsilon _{i}$ or $%
\varepsilon _{j}$. Due to these electron correlations, a resonant situation 
\begin{equation}
\mid \varepsilon _{i}-\varepsilon _{j}\mid \leq I_{ij}  \label{Eq.4}
\end{equation}
can be established for some short time interval $t_{I}$, where $I_{ij}$ $%
\varpropto $ exp$(-r_{ij}/a)$ is the overlap integral between sites $i$ and $%
j$. If the amplitude of the energy fluctuations is $E$ \TEXTsymbol{>}%
\TEXTsymbol{>} $I_{ij}$, this time interval can be approximated by

\begin{equation}
t_{I}\approx I_{ij}/(E\omega )  \label{Eq.5}
\end{equation}
The spreading out of the electron wave function between the resonant sites
obeying condition (4) takes a time $\tau $ of order of $\hbar /I_{ij}$. If
the energy fluctuations are fast compared to $\tau $, i.e., $\omega \tau $ 
\TEXTsymbol{>}\TEXTsymbol{>} 1, then the probability of the transition
occuring during one fluctuation period is $t_{I}/t$ and the transition rate
is

\begin{equation}
W\approx \omega t_{I}/\tau  \label{Eq.6}
\end{equation}
Using Eq.(5), one obtains

\begin{equation}
W\approx I_{ij}^{2}/(E\hbar )  \label{Eq.7}
\end{equation}

This formula shows that the frequency $\omega $ of the assisting hops is not
involved in the hopping rate for the current-carrying hops if the condition $%
\omega \tau $ \TEXTsymbol{>}\TEXTsymbol{>} 1 is fulfilled, i.e., if the
assisting hops are much faster than the current-carrying hop. This condition
is the main requirement of all SC models.

Above we considered the situation in which the current-carrying pair of
sites $i$ and $j$ already has one electron. In order to satisfy this
condition, one should take into account the energy distribution function of
the electrons $f(\varepsilon )$. Therefore, the phononless electron
transition rate $\nu _{ij}$ from the site $i$ to the site $j$ has the form

\begin{equation}
\nu _{ij}=Wf(\varepsilon _{i})[1-f(\varepsilon _{j})]  \label{Eq.8}
\end{equation}
where in the Ohmic regime, $f(\varepsilon )$ should be replaced by the Fermi
equilibrium distribution [8]. This leads to the expression

\begin{equation}
\nu _{ij}=\nu _{0}\exp \{-2rij/a\}\exp \{-\varepsilon _{ij}/kT\}
\label{Eq.9}
\end{equation}
for arbitrary positions of $\varepsilon _{i}$ and $\varepsilon _{j}$ with
respect to the Fermi level $\mu $. Here 
\begin{equation}
\varepsilon _{ij}=(1/2)\{\varepsilon _{i}-\varepsilon _{j}+\mid \varepsilon
_{i}-\mu \mid +\mid \varepsilon _{j}-\mu \mid \}  \label{Eq.10}
\end{equation}

It is worth noting that our expression for $\varepsilon _{ij}$ differs
slightly from the conventional one considered in hopping conduction, not
being determined by the phonon characteristics.

In order to calculate the exponential terms in the variable-range hopping
(VRH) conductivity using the phononless transition rates determined by
Eqs.(9) and (10), one can now repeat the routine derivation described in
detail in handbooks, e.g. Ref. [8]. This derivation automatically leads to
the dependences (1) and (2) [8]. All phonon-dependent factors in our
approach can appear only in terms of the quantity $\omega $, which is not
present in the final result for the hopping rate (Eq.(9)). The suggested
model of electron transport joins together a resonant tunneling mechanism of
band conductivity with the necessity of finding an electron in the initial
state and an empty final state, which is peculiar for the variable-range
hopping conductivity and results in the temperature dependence of this
band-like electron motion. It is therefore not surprising that experiments
[3,4] gives evidence for the dependences (1) and (2) in the phononless VRH
regime. We call our mechanism Variable-Range Resonant Tunneling.

At the present stage, it is not possible to determine by comparison with
experiment which particular phononless transport mechanism, - that of Ref.6
or the one suggested here - is responsible for the observed effects. Both
mechanisms lead to the same temperature dependences of the conductivity.
Another problem is to find the conditions for which phononless hopping
becomes more efficient than phonon-assisted hopping. The rarity of the
observations of the phononless VRH justifies, in most cases, the suggestion
that electron-phonon scattering is responsible for VRH transport. However,
in the vicinity of the metal-insulator transition, quantum effects should
become pronounced [9], and under such conditions, the phononless mechanism
can become dominant. Indeed, recent observations of phononless VRH [4] have
been made on the system close to the metal-insulator transition.

In conclusions, one can claim the following. A novel transport mechanism is
suggested for conduction in disordered systems with localized electrons
interacting via their Coulomb potentials. It is based on the resonant
single-electron tunneling between current-carrying sites brought into
resonance by fast electron hops in their surrounding. These assisting hops
can be due to electron-phonon scattering. If the typical rate of these
transitions is much higher than that of the current-carrying hops, the
former does not influence the result for the temperature dependence of VRH
conductivity.

We thank M. Pollak and A. M\"{o}bius for valuable discussions, in clarifying
the results of Ref.6 and Ref.10. Financial support of the
Deutscheforschungsgemeinschaft via SFB 383 and of the Fonds der Chemischen
Industrie is gratefully acknowledged.\medskip \newpage

\bigskip {\large References}

\bigskip \lbrack 1] G. Ebert, K. von Klitzing, C. Probst, E. Schubert, K.
Ploog, and G. Weimann, Solid State Commun. \textbf{45}, 625 (1983).

[2] A. Briggs, Y. Guldner, J.P. Vieren, M. Voos, J.P. Hirtz, and M. Razeghi,
Phys. Rev. B \textbf{27}, 6549 (1983).

[3] W. Mason, S.V. Kravchenko, G.E. Bowker, and J.E. Furneaux, Phys. Rev. B 
\textbf{52}, 7857 (1995).

[4] S.I. Khondaker, I.S. Shlimak, J.T. Nicholls, M. Pepper, and D.A.
Ritchie, in Proceedings of the 24th International Conference on the Physics
of Semiconductors, Jerusalem , 1998, edited by M. Heiblum and E. Cohen
(World Scientific, Singapore, 1998), in press; Phys. Rev. B, in press.

[5] I.L. Aleiner, D.G. Polyakov, and B.I. Shklovskii, in Proceedings of the
22nd International Conference on the Physics of Semiconductors, Vancouver,
1994, edited by D.S. Lockwood (World Scientific, Singapore, 1994), p. 787.

[6] L. Fleishman, D.C. Licciardello, and P.W. Anderson, Phys. Rev. Lett. 
\textbf{40}, 1340 (1978).

[7] M.L. Knotek and M. Pollak, J. Non-Cryst. Solids \textbf{8-10}, 505
(1972); Phys. Rev. B\textbf{\ 9}, 644 (1974).

[8] B.I. Shklovskii and A.L. Efros, ''Electronic Properties of Doped
Semiconductors'' (Springer, Berlin, 1984) p. 251.

[9] J. Talamantes, M. Pollak, and L. Elam, Europhys. Lett. \textbf{3}5, 511
(1996).

[10] J. Talamantes and A. M\"{o}bius, phys. stat. solidi (b) \textbf{205},
45 (1998).

[11] A. Perez-Garrido, M. Ortuno, E. Cuevas, J. Ruiz, and M. Pollak, Phys.
Rev. B \textbf{55}, R8630 (1997).

[12] N.V. Agrinskaya and V.I. Kozub, Semiconductors \textbf{32}, 631 (1998).

\end{document}